\documentclass[sigconf,nonacm]{acmart}

\AtBeginDocument{%
  \providecommand\BibTeX{{%
    \normalfont B\kern-0.5em{\scshape i\kern-0.25em b}\kern-0.8em\TeX}}}

\copyrightyear{2022}
\acmYear{2022}
\setcopyright{acmlicensed}\acmConference[FIRE '22]{Forum for Information Retrieval Evaluation}{December 9--13, 2022}{Kolkata, India}
\acmBooktitle{Forum for Information Retrieval Evaluation (FIRE '22), December 9--13, 2022, Kolkata, India}
\acmPrice{15.00}
\acmDOI{10.1145/3574318.3574339}
\acmISBN{979-8-4007-0023-1/22/12}

\newcommand\blfootnote[1]{%
  \begingroup
  \renewcommand\thefootnote{}\footnote{#1}%
  \addtocounter{footnote}{-1}%
  \endgroup
}

\usepackage{multirow}
\usepackage{soul}
\setstcolor{red}
\usepackage{graphicx}
\begin{document}

\title{Evaluating Impact of Social Media Posts by Executives on Stock Prices}

\author{Anubhav Sarkar}
\authornote{Both authors contributed equally to this research.}
\email{sarkaranubhav2001@gmail.com}
\orcid{0000-0003-3422-4232}

\author{Swagata Chakraborty}
\authornotemark[1]
\email{swagatac652@gmail.com}
\orcid{0000-0002-0216-7408}
\affiliation{%
  \institution{St. Xavier's College (Autonomous)}
  \city{Kolkata}
  \country{India}
}

\author{Sohom Ghosh}
\email{sohom1ghosh@gmail.com}
\orcid{0000-0002-4113-0958}
\author{Sudip Kumar Naskar}
\email{sudip.naskar@gmail.com}
\orcid{0000-0003-1588-4665}
\affiliation{%
  \institution{Jadavpur University}
  \city{Kolkata}
  \country{India}
}








\renewcommand{\shortauthors}{Sarkar and Chakraborty, et al.}

\begin{abstract}
Predicting stock market movements has always been of great interest to investors and an active area of research. Research has proven that popularity of products is highly influenced by what people talk about. Social media like Twitter, Reddit have become hotspots of such influences. This paper investigates the impact of social media posts on close price prediction of stocks using Twitter and Reddit posts. Our objective is to integrate sentiment of social media data with historical stock data and study its effect on closing prices using time series models. We carried out rigorous experiments and deep analysis using multiple deep learning based models on different datasets to study the influence of posts by executives and general people on the close price. Experimental results on multiple stocks (Apple and Tesla) and decentralised currencies (Bitcoin and Ethereum) consistently show improvements in prediction on including social media data and greater improvements on including executive posts. 
\end{abstract}





\maketitle

\section{Introduction}

\begin{figure}
\label{fig:intro}
\includegraphics[width=0.5\textwidth]{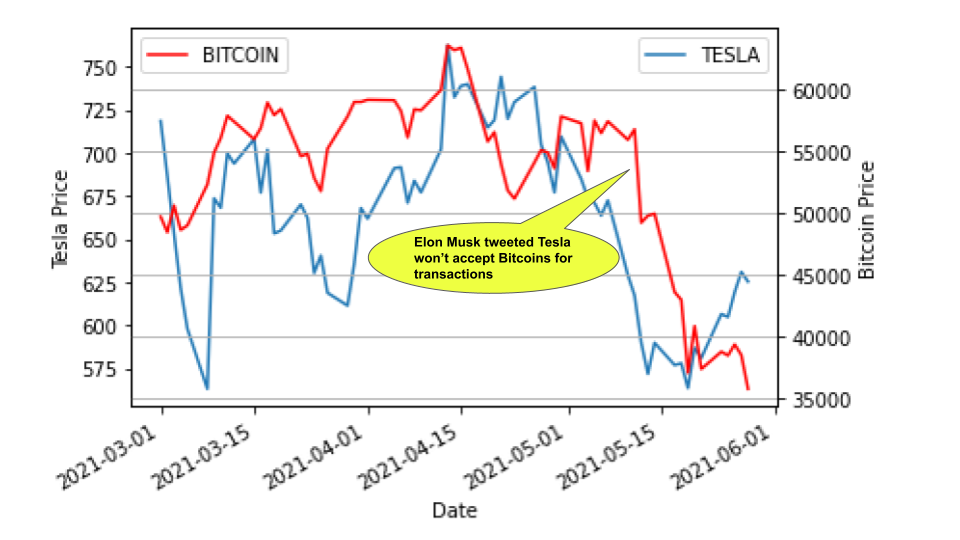}
\caption{Elon Musk's tweets and its effect on stock prices}
\centering
\end{figure}

\blfootnote{©Authors 2022. This is the author's version of the work. It is posted here for your personal use. Not for redistribution. The definitive version was published in the ACM proceedings of the 14\textsuperscript{th} meeting of Forum for Information Retrieval Evaluation (FIRE-2022), \url{https://doi.org/10.1145/3574318.3574339}}

Real world outcomes are highly influenced by the opinion of people. Social media has become the top priority platform for people to share their opinion about products, services, movies, stocks etc. These opinions influence others' decisions and thought processes. Research \cite{7-leskovec-dynamics-2007} has proved that marketing and popularity of a product or stock is highly influenced by what the society and its people think and talk about it. This has given rise to `meme stocks'. The world has witnessed how Elon Musk changing his Twitter bio to \texttt{\textbf{`\#bitcoin'}} caused a hike in the  price  of  bitcoin\footnote{\url{https://www.blockchainresearchlab.org/2021/02/08/the-musk-effect-how-elon-musks-tweets-affect-the-cryptocurrency-market/}, accessed on: 28\textsuperscript{th} June, 2022}. In fact, Elon Musk's decision to buy \$1.5 Billion of Bitcoin also caused the currency value to become sky high\footnote{\url{https://www.bbc.com/news/business-55939972}, accessed on: 28\textsuperscript{th} June, 2022}. This indicates to an underlying fact that the opinions of executives can bring changes in the real world.  
Motivated by this incident, we try to find answers to the following three research questions.
\begin{itemize}
\item \textbf{RQ-1:} Does social media have any influence on close price movements of stocks over a longer period of time?
\item \textbf{RQ-2:} Do opinions of executives on Twitter have greater influence on closing price of stocks than that of general people?
\item \textbf{RQ-3:} How does Reddit fare compared to Twitter with respect to the task of close price prediction?
\end{itemize}

Initially, researchers believed that publicly available historical stock data is the only factor affecting the next day stock price. However, gradually people realized the power of social media and witnessed how opinion of executive people were affecting the stock market movements.

We designed our first experiment to validate the hypothesis that opinions expressed on social media have a deep influence on the close price of stocks and decentralised currencies. Performance improvements were observed on integrating sentiment mined from social media data with historical stock data. The next set of experiments were performed to find out whether executive or general tweets have a deeper influence. Sentiment of general posts and sentiment of executive posts were separately integrated with historical stock data and different datasets were obtained on which we carried out the experiments. A better performance was witnessed on using sentiment of executive posts. Multiple experiments were performed on stocks (Tesla  and  Apple) and decentralised currencies (Bitcoin and Ethereum) to prove and validate the findings. Both Twitter and Reddit posts were considered for these experiments.

The paper is organized as follows. Section \ref{sec:related-works} gives an overview of the previous works in this field and positions our work. Section \ref{sec:data} deals with the various procedures of data collection, pre-processing and exploratory data analysis and gives us a summary of all the datasets that have been used in this work. Section \ref{sec:exp-setup} presents the different models and their architectures that have been used in the experiments. In section \ref{sec:experiments}, we discuss the experiments and present their results together with some analysis. Section \ref{sec:conclusion} answers the research questions through experimental results and concludes the paper mentioning some future directions of research. Finally, \ref{sec:limitations} and \ref{sec:ethics} throw light on limitations and ethical considerations, respectively.

We made the following contributions in the paper. 
\begin{itemize}
    \item We have validated that social media posts have an influence on close price movements. Subsequently, we have proposed how to use sentiments from social media to accurately predict the close prices. 
    \item We have shown that opinions of the executives matter more than opinions of the crowd in predicting the close price movements.
    \item 
    We showed that Reddit shows a similar trend like Twitter, however, Twitter is more effective in this task than Reddit.
\end{itemize}
\textbf{Reproducibility:}  Our code has been open-sourced\footnote{{\url{https://github.com/datagodno/Evaluating-Impact-of-Social-Media-Posts-by-Executives-on-Stock-Price}}} so that researchers can leverage it for future research. 
To ensure reproducibility of our results we have released a dataset comprising of ids of social media posts which were used in this research. To comply with the terms and conditions of Twitter and Reddit, we could not share the text content of the social media posts.

To the best of our knowledge, we are the first to extensively study the effects of sentiments of tweets and Reddit posts on stock prices of listed companies (Apple, Tesla) as well as decentralised currencies (Bitcoin, Ethereum).
 
\section{Related Works}
\label{sec:related-works}
Social media has become an integral part of our lives. We tend to express our thoughts and opinions by posting them on social media platforms like Twitter, Reddit, etc. \citet{7-leskovec-dynamics-2007} gathered information on how people converse regarding particular products and proved that it can be helpful in designing marketing and advertising strategies. \citet{bollen2011twitter} did one of the pioneering work by aggregating moods from Twitter using and assessing how these moods correlated with one of the market index (Dow Jones Industrial Average) over time. They used OpinionFinder and  Google-Profile of Mood States for the same. Traditionally, close price prediction of stocks were used to be done using methods like moving average, auto-regressive integrated moving average and so on. Presently, machine learning based algorithms have outperformed the traditional methods \cite{adebiyi2014comparison}. \citet{1-VIJH2020599} used Random Forest and Artificial Neural Networks for close price prediction. They collected historical data of five companies from Yahoo Finance for training these models. Using various performance metrics, they have proved that the Artificial Neural Network works better than Random Forest. However, they did not take into account the prevailing sentiments.   

\citet{mao-correlating-sp500-2012} established that the number of daily tweets mentioning `S\&P 500' was correlated with its daily closing price and absolute change in price. They further proved this correlation is more for industries like Finance, Energy, Materials and Healthcare. Subsequently, they showed that the daily traded volume and absolute change in price of Apple Inc.'s stock was positively correlated with the number of daily tweets relating to Apple. Similarly, \citet{sprenger2014tweets} showed how the positive sentiments and volumes of tweets were related to higher returns.  These tweets were posted between 1\textsuperscript{st} January and 30\textsuperscript{th} June 2010. 

\citet{lee2015role} studied how the social media posts of corporates relating to product recall limited the harm on their firm's reputation. \citet{5-pagolu-sentiment-2016} have successfully established a correlation between stock data and Twitter data only for Microsoft stock. But, they have not explored the same for other stocks and platforms like Reddit. \citet{6-asur-predicting-2016} used social media content to predict real-world outcomes like forecasting box-office revenues for movies.

\citet{eli-accr-51865-2017} focused on aggregated opinions of the people in general which is commonly referred to as wisdom of crowd. They proved that tweets of individuals could be used to forecast the earnings of an organization. On other side, \citet{elliott2018negative} and \citet{chen2019emergence} emphasized on the importance of tweets by executives.
\citet{jung-accr-51906-2017} concluded that firms tweeted less regarding their financial when their earnings were poor. They further studied how tweets related to bad earnings tarnished organizations' images through media coverage.
\citet{2-Jermann2017PredictingSM} used sentiment of executive tweets in predicting stock prices while ignoring that of the general people. Similarly, \citet{elliott2018negative} studied how tweets from CEOs after negative earnings helped in retaining investors. They concluded that CEOs who bonded with investors over Twitter gained were consider more trustworthy by the investors. \citet{chen2019emergence}  also presented similar findings. They studied how social media usage by executives of reputed firms impacted their stock prices and information environment. They further trained several machine learning models to classify executive tweets into three classes: company-related news announcement, work-related day-to-day activities  and unrelated to-work (i.e. personal posts). However, they used a static list of negative words to assess the negativity of tweets. \citet{seaton2019tweets} established that CEOs use social media platforms like Twitter to manipulate the investors. \citet{crowley2021executive} studied how the markets reacts to the tweets by executives and their firms during crucial business events. 

Lately, \citet{4-deshmukh2021stock} used stock data of multiple companies and performed sentiment analysis of tweets using Vader \cite{vader-Hutto_Gilbert_2014} to predict close price. \citet{chen2021opinion} presented an overview of various finance related opinion and argument mining techniques which are applicable on various sources like annual reports, earnings conference call, speeches, etc.

\citet{xu-cohen-2018-stock} proposed a neural-based model called StockNet for predicting rise or fall of various stock prices across nine industries. In addition to historical stock prices, this model used tweets for predicting the direction of movement of future stock prices.

\citet{chen-2021-www-eval} discussed how they used bi-directional Gated Recurrent Units \cite{cho-etal-2014-learning} along with BERT \cite{devlin-etal-2019-bert} and Convoluted Neural Networks \cite{kim-2014-convolutional} for distinguishing social media posts of amateur from expert investment professionals. They further proposed two metrics, maximum possible profit (MPP) and maximum loss (ML) for quantitatively measuring the quality of these posts. Lastly, they released the Investor’s ClaimRationale Dataset and proposed two tasks relating to rationale detection and claim-rationale inference.

\citet{8-seroyizhkosentiment-2022} integrated sentiment of Bitcoin based on Reddit posts with Bitcoin stock data but did not achieve much improvements. They concluded that integration of social media information in the form of sentiment is still an open research.

Recently, \citet{sawhney-etal-2022-cryptocurrency} presented CryptoBubbles, a novel task of detecting market bubbles relating to crypto-currencies. They curated the dataset from Reddit and Twitter posts which related to crypto-currencies and meme stocks. Finally, they proposed a Multi Bubble Hyperbolic Network for solving this task.

\section{Data}
\label{sec:data}
\subsection{Data Collection}
This section discusses how data was collected from three different sources. The procedures have been discussed below:

\subsubsection{Twitter Data}
Using snscrape\footnote{\url{https://github.com/JustAnotherArchivist/snscrape}, accessed on: 30\textsuperscript{th} June, 2022}, tweets about specific stocks were scrapped using their stock tickers like \textbf{`TSLA'} (for Tesla), \textbf{`AAPL'} (for Apple), \textbf{`BTC'} (for Bitcoin), and \textbf{`ETH'} (for Ethereum). We refer to this scraped tweet dataset as dataset \texttt{\textbf{T}}. This  dataset  contains features like date, username, and tweet of both executive and general people from 1\textsuperscript{st} January 2017 to 6\textsuperscript{th} May 2022.
A list of 122 executive Twitter handles was  obtained from Forbes\footnote{\url{https://www.forbes.com/sites/alapshah/2017/11/16/the-100-best-twitter-accounts-for-finance/?sh=783b0017ea0a}, accessed on: 30\textsuperscript{th} June, 2022}. This list includes notable people like Elon Musk, Warren Buffett, etc. From  dataset \texttt{\textbf{T}},  tweets  of  these  executives were separated. Thus, two datasets were obtained, referred to as Dataset \texttt{\textbf{E}} and Dataset \texttt{\textbf{G}}, for executive and general tweets, respectively. 

\subsubsection{Reddit Data}
Using pushshift.io\footnote{\url{https://github.com/pushshift/api}, accessed on: 30\textsuperscript{th} June, 2022} Reddit API, posts on particular subreddits were scrapped. The dataset contains features like upvotes, date, posts and the subreddit. It is referred to as dataset \texttt{\textbf{R}}. According to Investopedia\footnote{\url{https://www.investopedia.com/reddit-top-investing-and-trading-communities-5189322}, accessed on: 30\textsuperscript{th} June, 2022}, there are subreddits that can influence the stock market. These subreddits include \textit{`r/cryptocurrency'}, \textit{`r/investing\_discussion'}, \textit{`r/robinhood'}, \textit{`r/pennystocks'}, \textit{`r/investing'}, and \textit{`r/stock'}. Posts containing  these  executive subreddits  were scrapped to form a dataset, referred to as Dataset \texttt{\textbf{E\textsubscript{r}}}. Rest of the Tesla stock specific subreddits (`\textit{tsla', `TSLAtalk', `teslainvestorsclub', `TSLALounge', `TSLAsexy', `Tesla\_Stock' and `tslaq'}) 
were considered as general, and scrapped to form a dataset referred to as Dataset \texttt{\textbf{G\textsubscript{r}}}.

\subsubsection{Historical Stock Data}
Using Yahoo Finance\footnote{\url{https://finance.yahoo.com/}, accessed on: 30\textsuperscript{th} June, 2022}, we obtained the historical stock data separately for each company stock or decentralised currency from 1\textsuperscript{st} January 2017 to 6\textsuperscript{th} May 2022. This dataset contains the features -- `open': the share price of a single stock at the start of the day, `high': the highest price at which the stock was sold on that day, `low': the lowest price the stock was sold on that day, `volume' : total number of shares that were sold or bought on that day, and `close': the closing price of a single stock on that day.  Our objective was to build a model that can predict the `close price shifted': the close price of the next day. This dataset is referred to as Dataset \texttt{\textbf{Y}}. 

\subsubsection*{Notations} Table \ref{tab:notation} presents a list of notations used in this paper and their descriptions.

\subsection{Exploratory Data Analysis}
\begin{figure*}
\caption{Number of executive and general tweets per day}
\label{fig:eda-tweet-day}
\includegraphics[width=1.0\textwidth]{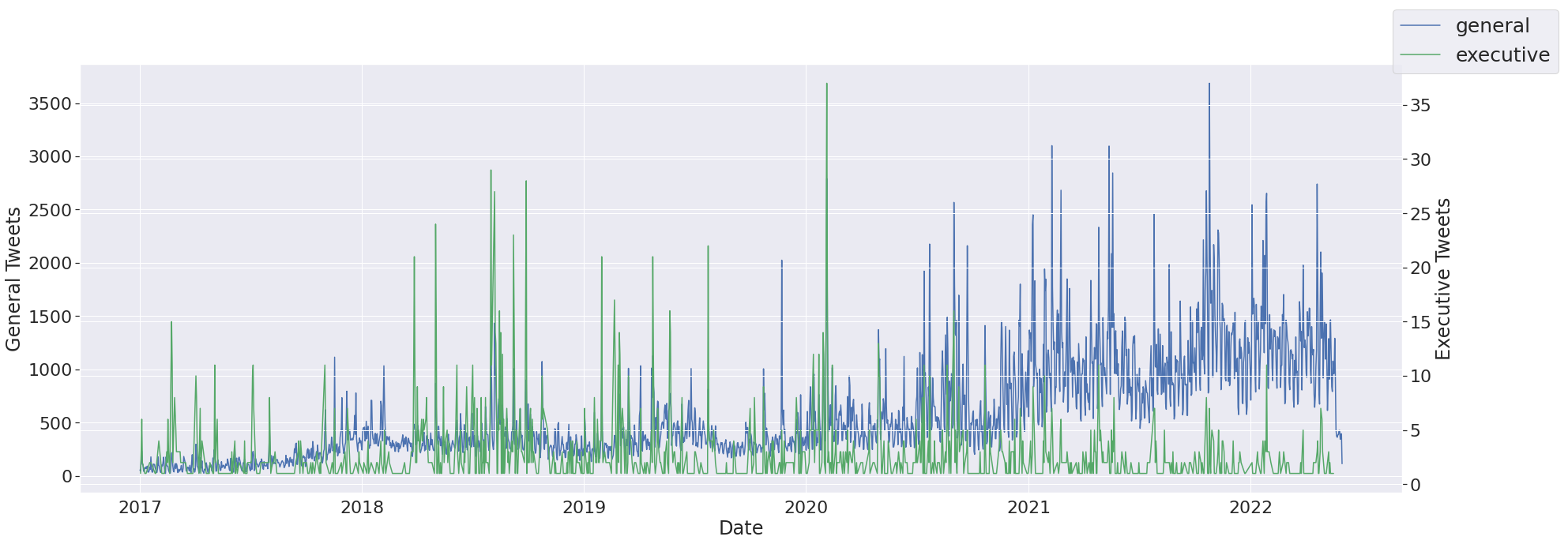}
\centering
\end{figure*}

Tweets relating to Tesla, Apple, Bitcoin, and Ethereum were collected. 
The collected tweets were made by some executives and largely by general people. 
The total number of executive posts that were collected was 4,470 and the total number of non-executive tweets that were collected was 1,207,144. Since, tweets made by general people overshadow executive tweets, we perform under-sampling of the majority class. We limit the general posts to approximately 19,000 tweets for every stock. In case of Yahoo Finance, data were scrapped from 1\textsuperscript{st} January, 2017 to 5\textsuperscript{th} May, 2022. However, no data was available in the weekends for different stocks and no data was available for Ethereum for the first 10 months of 2017 for Ethereum. This resulted in different counts of days with closing prices. We present the stock-wise statistics in Table \ref{tab:ex-non-stck}. Figure \ref{fig:eda-tweet-day}, shows tweets made by executives and generals per day. The green line corresponds to executive posts and the blue line corresponds to general posts in Figure \ref{fig:eda-tweet-day}. 

\begin{table*}
\centering
\caption{Notations and descriptions of the corresponding datasets}
\label{tab:notation}
\begin{tabular}{ll}
\toprule
\textbf{Notation}                                                & \textbf{Data Description} \\
\midrule
\textbf{X (T)}                                                &   Tweets relating to stock X\\ 
\textbf{TSLA (R)}                                                    & Reddit posts relating to TSLA                     \\
\texttt{\textbf{Y}}+\texttt{\textbf{T}}$_{vader}$                                                        &   Closing prices \& Vader based sentiment scores of all tweets \\
\texttt{\textbf{Y}}+\texttt{\textbf{T}}$_{finbert}$                                                      &   Closing prices \& FinBERT based sentiment scores of all tweets                 \\
\texttt{\textbf{Y+G}}                                                              &   Closing prices \& FinBERT based sentiment scores of general tweets                  \\ 
\texttt{\textbf{Y+E}}                                                              &      Closing prices \& FinBERT based sentiment scores of executive tweets                 \\
\texttt{\textbf{Y+G\textsubscript{r}}}                                                             &  Closing prices \& FinBERT based sentiment scores of general reddit posts \\
\texttt{\textbf{Y+E\textsubscript{r}}}                                                              & Closing prices \& FinBERT based sentiment scores of executive reddit posts                    \\
\bottomrule
\end{tabular}
\end{table*}

\begin{table*}
\caption{Distribution of Social Media Posts}
\label{tab:ex-non-stck}
\begin{tabular}{lclrrrr}
\toprule
\textbf{Stock} & \textbf{\# Days with closing prices} & \textbf{Category} & \textbf{\# Posts} & \textbf{Reduced \# Posts} & \textbf{\# Days with Posts} & \textbf{\# Days with no Posts} \\
\midrule
\multirow{2}{*}{\textbf{TSLA (T)}}&\multirow{2}{*}{1,346} & \texttt{Executive} & 2,617 & NA & 769 & 577 \\ & & \texttt{General} & 4,82,375 & 19,164 & 1,346 & 0       \\ \hline
\multirow{2}{*}{\textbf{AAPL (T)}}&\multirow{2}{*}{1,346} & \texttt{Executive} & 260 & NA & 179 & 1,167 \\ & & \texttt{General} & 21,383 & 19,057 & 1,324 & 22     \\ \hline
\multirow{2}{*}{\textbf{BTC (T)}}&\multirow{2}{*}{1,894} & \texttt{Executive} & 303 & NA & 242 & 1,652\\ & & \texttt{General} & 5,38,442 & 19,022 & 1,894 & 0    \\ \hline
\multirow{2}{*}{\textbf{ETH (T)}}&\multirow{2}{*}{1,543} & \texttt{Executive} & 51 & NA & 46 & 1,497 \\ & & \texttt{General} & 1,53,362 & 19,091 & 1,535 & 8 \\ \hline
\multirow{2}{*}{\textbf{TSLA (R)}}&\multirow{2}{*}{952} & \texttt{Executive} & 1,239 & NA & 98 & 854 \\ & & \texttt{General} & 11,582 & NA & 558 & 394\\
\bottomrule
\end{tabular}
\end{table*}



\
subsection{Data Pre-processing}
The tweets (i.e., datasets \texttt{\textbf{E}} and \texttt{\textbf{G}}) as well as Reddit posts were subjected to similar pre-processing steps. 
While extracting tweets, retweeted tweets were also considered  as unique (i.e., separate) tweets. To avoid duplicacy, the duplicate tweets for every user were dropped. Every day thousands of people tweets relating to a given stock. To understand the sentiment associated with a given stock for a particular day, we extracted sentiments from all the tweets mentioning the stock in that day and averaged them. 
We used Vader \cite{vader-Hutto_Gilbert_2014} and a pre-trained language model FinBERT \cite{araci2019finbert} for obtaining scores corresponding to three different types of sentiment -- Positive, Negative, and Neutral. For  FinBERT \cite{araci2019finbert}, these scores were normalized using the softmax function. For each day, we considered the average sentiment scores of all the tweets on that day. The day wise aggregated sentiment scores were aligned to the dataset \texttt{\textbf{Y}}.
Dates for which no tweets were available, the sentiment scores of those dates were imputed using Cubic Spline Interpolation\footnote{\url{https://pythonnumericalmethods.berkeley.edu/notebooks/chapter17.03-Cubic-Spline-Interpolation.html}, accessed on: 5\textsuperscript{th} July, 2022} technique for both executive and non-executive missing sentiment scores. A simple average method was not  chosen since that approximation would be biased and much different from the actual value. 
Large language models like FinBERT \cite{araci2019finbert} needs high computational resources for training and scoring.
Due to computational constraints, we considered a random sample of around 19,000 general tweets. Table \ref{tab:ex-non-stck}, shows the number of days with no posts per stock for executives as well as general people. Unlike the decentralised currencies (Bitcoin and Ethereum), closing prices of listed companies (Apple and Tesla) are not available during the weekends.
Subsequently, based on the availability of data we had to adjust the starting date for our analysis. Thus, the number of days with closing prices is different for different stocks. Whereever we did not need to under-sample the number of posts, we mark it as NA (i.e. Not Applicable).

To predict close price more accurately, researchers \cite{1-VIJH2020599} have introduced new variables which are derived from existing variables. In the present work we considered the exponentially weighted moving average (\textit{ewma}) for the closing price and the sentiment scores for 3, 7, 14 and 30 days 
The intuition behind using four different ewma values is to cover all sudden and long-term changes to the price and sentiment of the stock. The derived dataset thus has more features: closing price, ewma closing prices, volume, open price, high, low, sentiment scores (corresponding to positive, negative and neutral classes) and ewma of sentiment scores. The goal of this work is to predict the close price of the next day,  therefore a close price shift was added as the target variable for prediction. The dataset after being normalised using Standard Scaler was divided into training and test sets maintaining a ratio of 80\% to 20\%. The initial date range corresponding to the training data and the later date range corresponding to the test data are mentioned in Table \ref{tab:train-test}. The starting dates are different for different stocks since missing posts right at the beginning of the date range could not be imputed. Imputation only works when data is missing in between the date range.

\begin{table}
\caption{Train and Test splits}
\label{tab:train-test}
\begin{tabular}{lcccc}
\toprule
\multicolumn{1}{c}{\textbf{Stock}} & \textbf{Category} & \textbf{Start} & \textbf{End} \\
\midrule
\multirow{2}{*}{\textbf{TSLA (\texttt{\textbf{T}})}} & Train & 4\textsuperscript{th}, Jan 2017  & 14\textsuperscript{th}, Apr 2021 \\ 
                                   & Test  & 15\textsuperscript{th} Apr, 2021 & 5\textsuperscript{th} May, 2022 \\ \hline
\multirow{2}{*}{\textbf{AAPL (\texttt{\textbf{T}})}} & Train & 4\textsuperscript{th} Jan, 2017 & 27\textsuperscript{th} Apr, 2021 \\ 
                                   & Test  & 28\textsuperscript{th} Apr, 2021 & 5\textsuperscript{rd} May, 2022  \\ \hline
\multirow{2}{*}{\textbf{BTC (\texttt{\textbf{T}})}}  & Train & 2\textsuperscript{nd} Mar, 2017 & 22\textsuperscript{rd} Apr, 2021 \\
                                   & Test  & 23\textsuperscript{th} Apr, 2021 & 6\textsuperscript{th} May, 2022\\ \hline
\multirow{2}{*}{\textbf{ETH (\texttt{\textbf{T}})}}  & Train & 16\textsuperscript{th} Feb, 2018 & 2\textsuperscript{nd} Jul, 2021  \\
                                   & Test  & 3\textsuperscript{rd} Jul, 2021 & 6\textsuperscript{th} May, 2022   \\ \hline
\multirow{2}{*}{\textbf{TSLA (\texttt{\textbf{R}})}} & Train & 30\textsuperscript{th} Jul, 2018 & 4\textsuperscript{th} Aug, 2021  \\
                                   & Test  & 5\textsuperscript{th} Aug, 2021 & 5\textsuperscript{th} May, 2022  \\
\bottomrule                                   
\end{tabular}
\end{table}

\section{Experimental Setup}
\label{sec:exp-setup}
This section discusses the architecture and setup of the various models that we used in our experiments. Due to the sequential nature of the data, we primarily used sequence-based models such as Recurrent Neural Networks (\textbf{RNN}) \cite{rnn}, Gated Recurrent Unit (\textbf{GRU}) \cite{gru}, Long Short Term Memory (\textbf{LSTM}) \cite{lstm} and Auto Encoders (\textbf{AE}) \cite{autoencoder}.
We initiated by creating an \textbf{RNN} model. This was initialised by Glorot Normal \cite{pmlr-v9-glorot10a} values. To generate the same random weights every time, the seed value was set to 42. It had three sequential RNN layers with a dropout \cite{JMLR:v15:srivastava14a} rate of 0.4 in each layer. The layers had 250, 200 and 150 neurons respectively.
Subsequently, an output dense layer of a single neuron with a linear activation function was added. Adam optimiser \cite{adam} with a learning rate of 0.0001 was used. Mean Squared Error was used as the loss function. The model was run for 250 epochs with a batch size of 16 and a validation split of 0.1. Early stopping was performed with a patience value of 5 and the best weights were restored. 
Keeping everything else unaltered, we replaced the RNN layers by bi-directional RNN (\textbf{Bi-RNN}) \cite{bi-rnn} layers in the above experiment. We further repeated the same experiment by replacing the RNN layers with \textbf{GRU} \cite{gru}, bi-directional GRU (\textbf{bi-GRU}), \textbf{LSTM} \cite{lstm} and bi-directional LSTM (\textbf{bi-LSTM}) layers. 
Lastly, we trained an Auto Encoder model (\textbf{AE}) consisting of a bi-directional LSTM layer with 250 neurons, tanh activation function, and a drop out rate of 0.4. This layer was followed by another LSTM layer with 200 neurons, a repeat vector layer, another two LSTM layers with 200 and 250 neurons each and drop out rate of 0.4 and 0.3 respectively. Finally, we added a flatten layer with a dropout rate of 0.4 and a dense layer with linear activation function.

\subsection*{Performance Metrics}
We used MAE (Mean Absolute Error), RMSE (Root Mean Square Error), Adjusted R$^{2}$ (R$^{2}_{a}$) and MAPE (Mean Absolute Percentage Error) to evaluate the models. Among these metrics MAE, RMSE and MAPE are error metrics, i.e., the lower the value the better, while R$^{2}_{a}$ is an accuracy metric.

\section{Experiments and Results}
\label{sec:experiments}
To address the research questions, we carried out multiple experiments which are discussed in the following subsections. All of the experiments were performed in Google Colab with GPU.

\subsection{Effect of Social Media Sentiment on Prediction of Close Price (Experiment 1)}

\begin{figure*}
\caption{Close price prediction of Tesla with Y, Y+T$_{vader}$ and Y+T$_{finbert}$ datasets using LSTM}
\label{fig:with-out-senti}
\includegraphics[width=\textwidth]{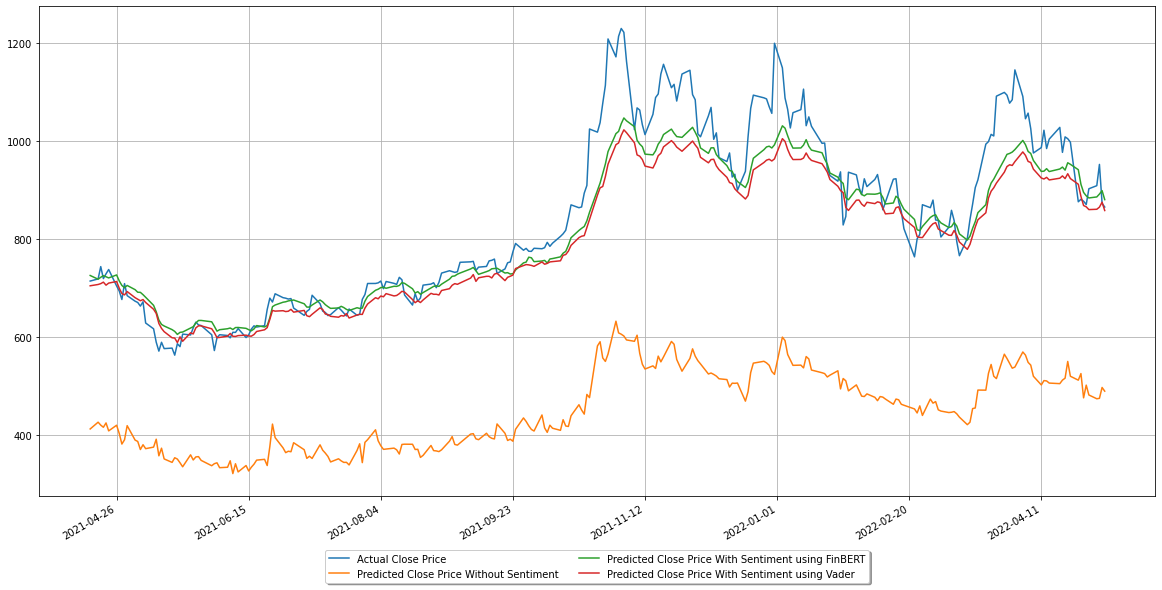}
\centering
\end{figure*}

This experiment was performed to answer RQ1, i.e., to investigate whether social media sentiment about a particular stock contributes to predicting its close price. For this study we chose the Tesla as the stock, twitter as the social media, and VADER \cite{vader-Hutto_Gilbert_2014} and FinBERT \cite{araci2019finbert} tools as the sentiment analysis models. We used the historical stock data of Tesla from Yahoo (\texttt{\textbf{Y}}), 
and sentiment scores obtained from VADER and FinBERT on tweets (\texttt{\textbf{T}}) about Tesla from Twitter. Sentiment analysis was performed on dataset \texttt{\textbf{T}} using two sentiment analysis models, VADER -- a rule based system, and FinBERT -- a pre-trained model built by finetuning the BERT language model in the finance domain for preforming sentiment analysis of financial text. After sentiment analysis, we obtained two different datasets, \texttt{\textbf{T}}$_{vader}$ and \texttt{\textbf{T}}$_{finbert}$. We obtained the scores corresponding to every type of sentiment (`positive', `negative' and `neutral'). 
These two datasets were merged according to dates with dataset \texttt{\textbf{Y}} giving rise to two datasets --  \texttt{\textbf{Y}}+\texttt{\textbf{T}}$_{vader}$ and \texttt{\textbf{Y}}+\texttt{\textbf{T}}$_{finbert}$, having 1,346 instances of 24 features each. These 24 features consist of 5 features from the original dataset \texttt{\textbf{Y}} (`open', `high', `low', `close' prices \& `volume' traded), 3 scores corresponding to sentiments (positive, negative \& neutral) and 16 derived features (i.e. exponentially weighted moving averages for 3, 7, 14 \& 30 days  of `close' price, positive, negative \& neutral sentiment scores).
The dependent variable (output) is the close price of the next day. Model trained on \texttt{\textbf{Y}} serves as our baseline model.
The LSTM model  was trained on these 3 datasets - \texttt{\textbf{Y}}, \texttt{\textbf{Y}}+\texttt{\textbf{T}}$_{vader}$ and \texttt{\textbf{Y}}+\texttt{\textbf{T}}$_{finbert}$, separately, and the results of these experiments are reported in Table \ref{tab:exp-1}. 
Results in Table \ref{tab:exp-1} shows significant improvement in performance across all evaluation metrics for both \texttt{\textbf{Y}}+\texttt{\textbf{T}}$_{vader}$ and \texttt{\textbf{Y}}+\texttt{\textbf{T}}$_{finbert}$ over the baseline model \texttt{\textbf{Y}}. This proves the phenomenal fact that the sentiment of social media data has immense influence in predicting the close price and it comprehensively answers RQ1. 
The experimental results further suggest that FinBERT provides much better performance with respect to this extrinsic evaluation, hence FinBERT is used for sentiment analysis in all the experiments henceforward. Figure \ref{fig:with-out-senti} pictorially presents the results of the prediction models. 
It shows that the curves obtained with the sentiment obtained using FinBERT (green) and the sentiment obtained using VADER (red) are much closer to the actual close price (blue) than the curve obtained without sentiment (orange). 

\begin{table}
\caption{Results of Experiment 1}
\label{tab:exp-1}
\begin{tabular}{ccccc}
\toprule
\textbf{Dataset}    & \textbf{MAE}     & \textbf{RMSE}    & \textbf{R$^{2}_{a}$}    & \textbf{MAPE  (\%)}   \\
\midrule
\textbf{Y}          & 393.195 & 405.845 & -4.434 & 46.193 \\
\textbf{Y+\texttt{\textbf{T}}$_{vader}$}   & 52.089          & 72.794          & 0.825      & 5.493 \\
\textbf{Y+\texttt{\textbf{T}}$_{finbert}$} & \textbf{42.186}  & \textbf{60.739}  & \textbf{0.878}  & \textbf{4.506} \\
\bottomrule
\end{tabular}
\end{table}

\subsection{Comparative Study of Models Predicting Close Price (Experiment 2)}


Using the \texttt{\textbf{Y+T}}$_{finbert}$ dataset, we experimented with  multiple models to find the best working model on this task. As the supremacy of neural networks over traditional methods like ARIMA \cite{adebiyi2014comparison} has been well-established for time series analysis, we tried various neural network-based architectures such as \textbf{RNN} \cite{12-rumelhart1985learning}, \textbf{GRU} \cite{gru}, \textbf{LSTM} \cite{lstm} and Auto-Encoder \cite{autoencoder}. Table \ref{tab:exp-2} presents the performance of these models on the \texttt{\textbf{Y+T}}$_{finbert}$ dataset. Among all these models, \textbf{GRU} provides the best working model across all evaluation metrics.
Hence, in all further experiments \textbf{GRU} is used as the close price prediction model. This model follows the architecture of the \textbf{GRU} Model mentioned in section \ref{sec:exp-setup}.

\begin{table}
\caption{Results of Experiment 2 on the \texttt{\textbf{Y+T}}$_{finbert}$ dataset}
\label{tab:exp-2}
\begin{tabular}{ccccc}
\toprule
\textbf{Model}   & \textbf{MAE}     & \textbf{RMSE}    & \textbf{R$^{2}_{a}$}   & \textbf{MAPE  (\%)}   \\
\midrule
\textbf{RNN}     & 184.194 & 233.086 & -0.889 & 19.341 \\
\textbf{Bi-RNN}  & 165.268  & 203.304  & -0.437 & 18.141  \\
\textbf{GRU}     & \textbf{26.688}  & \textbf{36.388}  & \textbf{0.953} & \textbf{3.061}  \\
\textbf{Bi-GRU}  & 30.509           & 42.166           & 0.938          & 3.478           \\
\textbf{LSTM}    & 78.982          & 105.288           & 0.614          &8.542           \\
\textbf{Bi-LSTM} & 87.046           & 107.885           & 0.595          &9.399           \\
\textbf{AE}      & 57.885           & 79.189          & 0.781          & 6.155          \\
\bottomrule
\end{tabular}
\end{table}


\subsection{Influence of  Executive Posts vs General Posts on Closing Prices (Experiment 3)}

This set of experiments were carried out to answer RQ2, i.e., whether opinions of executives have greater influence on closing price than that of general people. Firstly, we use sentiment scores of tweets about Tesla from Twitter and historical stock data of Tesla from Yahoo. Two datasets were used, dataset \texttt{\textbf{E}} and dataset \texttt{\textbf{G}}, for executive and general posts, respectively. These datasets were subjected to sentiment analysis. Then they were merged according to dates with dataset \texttt{\textbf{Y}}. Thus, we had two new datasets, \texttt{\textbf{Y+G}} and \texttt{\textbf{Y+E}} having 1,346 instances of 24 features each. The output is a single feature, i.e., the next day close price. The \textbf{GRU} Model was trained on these datasets individually and next day close price was predicted. We refer to this as Experiment 3.1. We extended this experiment and replicated the same experiment by using tweets and stock prices of Apple instead of Tesla. We refer to this as Experiment 3.2.

We further extended our experiments to two unlisted decentralised currencies:  Bitcoin and Ethereum. Unlike Tesla and Apple, these currencies do not depend on supply chain related factors like the availability of raw materials. Experiments 3.3 and 3.4 were carried out on Bitcoin and Ethereum datasets, respectively, keeping the same experimental framework, i.e., using the sentiment of posts about those currencies from Twitter and  the corresponding historical stock data from Yahoo. 

Finally, to answer RQ3, i.e., whether the above findings obtained using tweets also hold for Reddit, we repeated the same experiment using the sentiment of posts about Tesla on Reddit. This experiment is referred to as Experiment 3.5. This gives us a more comprehensive view of the bigger picture across two different social media platforms. This experiment has not been repeated with Apple, Bitcoin or Ethereum because of the tedious process of data collection which keeps failing multiple times due to payload.

\begin{figure*}
\caption{Close price prediction of Tesla with Y+G and Y+E Datasets}
\label{fig:exe-vs-gen}
\includegraphics[width=\textwidth]{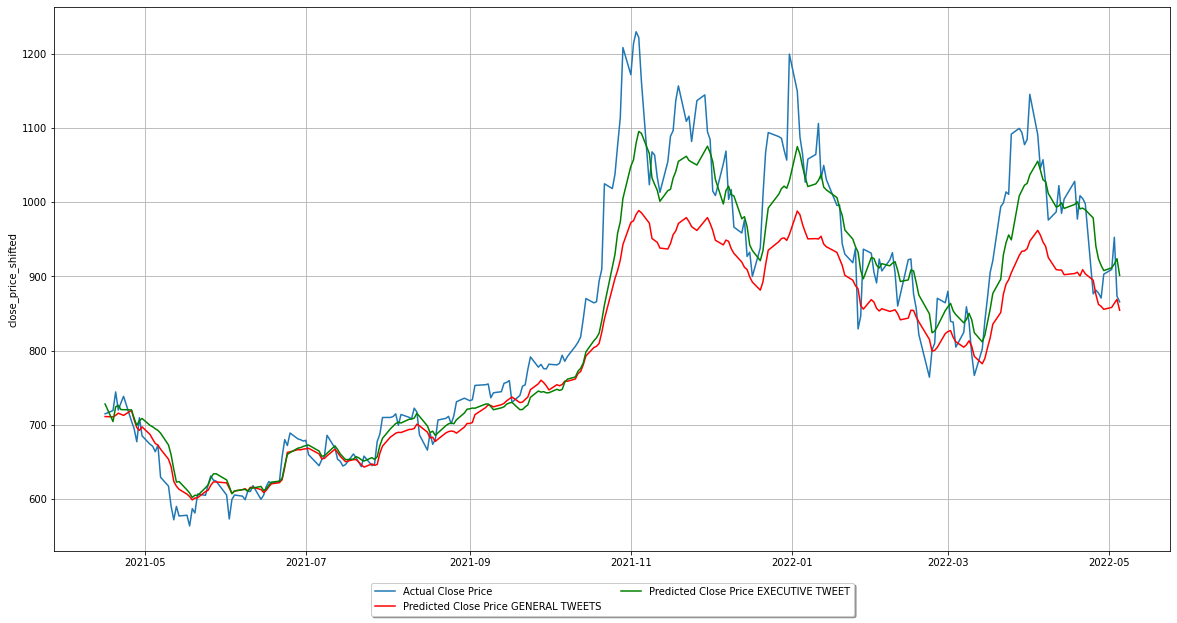}
\centering
\end{figure*}

Table \ref{tab:exp-4-5-6} reports the results of Experiments 3.1--3.5. Results of Experiments 3.1--3.5 clearly suggest that opinions of executives matter much more than opinions of general people in the close price prediction task since the \texttt{\textbf{Y+E}} and \texttt{\textbf{Y+E\textsubscript{r}}} datasets provide much better performance than the corresponding \texttt{\textbf{Y+G}} and \texttt{\textbf{Y+G\textsubscript{r}}} datasets respectively across all the evaluation metrics. Since the trend holds true across both Twitter and Reddit and for all the stocks and decentralised currencies considered, it proves that the finding is widespread and effective in multiple domains. Hence, we can concretely conclude that the influence of executive posts on close price is much more than general posts not only for different stocks but also for different decentralised currencies. Figure \ref{fig:exe-vs-gen}, plots the actual data of the close price (blue) for Tesla, the predicted close prices obtained with the sentiment of executive posts (green), and  general posts (red). It is evident from figure \ref{fig:exe-vs-gen} that the green curve is much closer to the blue curve than the red curve, i.e., executive opinions are much more effective than general opinions with respect to the close prediction task.

Overall, sentiments expressed by executive in Twitter gives the best performance. It is equally interesting to note that unlike Twitter, the difference in performance between the general and executive datasets is not significant for Reddit.


\begin{table}
\caption{Result of Experiments 3}
\label{tab:exp-4-5-6}
\resizebox{\columnwidth}{!}{
\begin{tabular}{llccccc}
\toprule
\multicolumn{1}{c}{\textbf{Exp}} & \multicolumn{1}{c}{\textbf{Stock}} & \textbf{Data} & \textbf{MAE}      & \textbf{RMSE}     & \textbf{R$^{2}_{a}$}   & \textbf{MAPE (\%)}   \\
\midrule
\multirow{2}{*}{\textbf{Exp 3.1}} & \multirow{2}{*}{\textbf{TSLA (T)}} & \texttt{\textbf{Y+G}} & 56.365 & 79.701 & 0.790 & 5.852 \\
                                 &                                    & \textbf{\texttt{\textbf{Y+E}}}  & \textbf{34.362}   & \textbf{48.261}   & \textbf{0.923}          & \textbf{3.817 }          \\ \hline
\multirow{2}{*}{\textbf{Exp 3.2}}  & \multirow{2}{*}{\textbf{AAPL (T)}}     & \texttt{\textbf{Y+G}}  & 4.700    & 5.922    & 0.859 & 2.932  \\
                                 &                                    & \textbf{\texttt{\textbf{Y+E}}}  & \textbf{3.075}             & \textbf{3.860}             & \textbf{0.940}          & \textbf{1.990}           \\ \hline
\multirow{2}{*}{\textbf{Exp 3.3}}  & \multirow{2}{*}{\textbf{BTC (T)}}               & \texttt{\textbf{Y+G}}  & 4681.737 & 5584.437 & 0.572 & 9.675 \\
                                 &                                    & \texttt{\textbf{Y+E}}  & \textbf{2842.190}          & \textbf{3679.708}          & \textbf{0.814}         & \textbf{5.830}          \\ \hline
\multirow{2}{*}{\textbf{Exp 3.4}}                                 & \multirow{2}{*}{\textbf{ETH (T)}}               & \texttt{\textbf{Y+G}}  & 315.075  & 407.849  & 0.641 & 8.663  \\
                                 &                                    & \texttt{\textbf{Y+E}}  & \textbf{278.507}           & \textbf{356.633}           & \textbf{0.725}          & \textbf{7.724}         \\ \hline
\multirow{2}{*}{\textbf{Exp 3.5}}                                 & \multirow{2}{*}{\textbf{TSLA (R)}}               & \texttt{\textbf{Y+G\textsubscript{r}}}  & 44.625 & 61.403 & 0.811 & 4.481  \\ 
                                 &                                    & \texttt{\textbf{Y+E\textsubscript{r}}}  &  \textbf{42.283}         &  \textbf{58.994}         &  \textbf{0.826}         &   \textbf{4.247}         \\                                 
\bottomrule                                 
\end{tabular}
}
\end{table}


\subsection{Comparative study of close price prediction with and without imputation (Experiment 4)}
This experiment was performed on all stocks and decentralised currencies using the datasets: \texttt{\textbf{Y+G}}, \texttt{\textbf{Y+E}}, \texttt{\textbf{Y+G\textsubscript{r}}} and \texttt{\textbf{Y+E\textsubscript{r}}}. The motivation behind this experiment is the observation that there is an abundance of tweets by general people and a scarcity of tweets by executives. To keep the research fair, we decided to equalise the number of tweets by general people and executives. We identified the dates on which there were no executive tweets. Tweets by general people were dropped for these dates. We performed Cubic Spline Interpolation for all the datasets. It resulted in an equal amount of data for the general and executive datasets. These datasets were used to train on the \textbf{GRU} model for predicting the close price of the next day. Table \ref{tab:exp-4}, shows the evaluation results for this experiment. 
We observe that for all the stocks, sentiments of tweets by executives are better predictors than that of the crowd.
However, in case of Reddit, the sentiments of general posts prove to be more effective than that of the executives. Since there is not much difference in our findings from experiments 3 and 4, we conclude that the imputation methodology we followed and the non-availability of executive posts for all days did not have much effect on the results.



\begin{table}
\caption{Result of Experiment 4}
\label{tab:exp-4}
\begin{tabular}{lccccc}
\toprule
\multicolumn{1}{c}{\textbf{Stock}} & \textbf{Data} & \textbf{MAE}    & \textbf{RMSE}   & \textbf{R$^{2}_{a}$}     & \textbf{MAPE  (\%)}   \\
\midrule
\multirow{2}{*}{\textbf{TSLA (\texttt{\textbf{T}})}} & \texttt{\textbf{Y+G}} & 59.621   & 83.765  & 0.768 & 6.174  \\
                                   & \texttt{\textbf{Y+E}}  & \textbf{34.362}          & \textbf{48.261}          & \textbf{0.923}            & \textbf{3.817}           \\ \hline
\multirow{2}{*}{\textbf{AAPL (\texttt{\textbf{T}})}}          & \texttt{\textbf{Y+G}}  & 3.899 & 5.004 & 0.899 & 2.459 \\
                                   & \texttt{\textbf{Y+E}}  & \textbf{3.071}          & \textbf{3.849}          & \textbf{0.940}          & \textbf{1.988}          \\  \hline
\multirow{2}{*}{\textbf{BTC (\texttt{\textbf{T}})}}           & \texttt{\textbf{Y+G}} &3676.006  &4631.177  &0.706  & 7.492  \\
                                   & \texttt{\textbf{Y+E}}  & \textbf{2842.190}        & \textbf{3679.708}        & \textbf{0.814}            & \textbf{5.830}          \\  \hline
\multirow{2}{*}{\textbf{ETH (\texttt{\textbf{T}})}}           & \texttt{\textbf{Y+G}} & 309.507  & 392.286  & 0.668 &8.557  \\
                                   & \texttt{\textbf{Y+E}}  & \textbf{278.507}         & \textbf{356.633}         & \textbf{0.725}            & \textbf{7.724}          \\  \hline
\multirow{2}{*}{\textbf{TSLA (\texttt{\textbf{R}})}}          & \texttt{\textbf{Y+G\textsubscript{r}}} & \textbf{46.953}   & \textbf{68.659}   & \textbf{0.844}  & \textbf{4.954}  \\
                                   & \texttt{\textbf{Y+E\textsubscript{r}}}  & 57.674         & 82.634        & 0.774          & 6.043 \\        
\bottomrule                                   
\end{tabular}
\end{table}

\section{Conclusions}
\label{sec:conclusion}
In this research we studied how the sentiment of social media posts by executives and people in general affect stock prices of two popular companies Apple and Telsa. We primarily considered Twitter and Reddit posts for this research. We extended our study by predicting prices of two popular decentralised currencies Bitcoin and Ethereum.
Our experiments successfully answer the research questions raised before. 

\textbf{RQ-1:} Social media data from both  Twitter and Reddit have a deep influence on close price movements. On integrating the sentiment of social media data, significant improvements were witnessed in close price prediction.

\textbf{RQ-2:}  Sentiment of tweets by executives have a deeper influence on the prediction of close price. This is because the executives have more impact on the society and the mass tends to have more faith in executives and are easily  influenced by the opinion of executive people. However, the effect of tweets by general people should not be considered unimportant. This is in supporting the claims made by Jermann \cite{2-Jermann2017PredictingSM}, Elliott et al. \cite{elliott2018negative} and Chen et al. \cite{chen2019emergence}.

\textbf{RQ-3:} Our findings using tweets also hold good for Reddit posts. 

This work has a lot of directions where further research could be performed. 
Instead of just using the sentiment of the tweets, we would like to use the entire textual content for predicting close prices.
A better way could be found to impute sentiment on days no tweets or posts are available.
If these models are trained on more granular data, users can leverage them for choosing winning stocks by utilising the stock price prediction made every minute.

\section{Limitations}
\label{sec:limitations}
The unavailability of executive posts has been a limitation of this work. It has been noticed that executives do not post much like the general people. Hence, general tweets and posts were found in abundance while executive posts were sparse. 
Moreover, we have considered only the tweets containing the aforesaid tickers. Other tweets about these companies which were not tagged with these tickers was not considered for our analysis. As mentioned in Table \ref{tab:train-test}, we needed to adjust the starting and ending dates slightly based on the availability of data. Furthermore, we did not consider features other than prices of stocks, traded volumes and sentiment scores.  Not all tweets are genuine. This study does not verify the authenticity of the tweets being used for analysis. 

\section{Disclaimer and Ethical Considerations}
\label{sec:ethics}
This research has been performed for academic purposes only. The authors declare that there are no commercial interests related to this. The opinions expressed here are that of the authors and not their affiliations.

We developed the dataset by collecting publicly available posts from Twitter and Reddit. We followed all the ethics and rules set by these organizations. Consent of individual users for using their posts was not necessary.


\bibliographystyle{ACM-Reference-Format}
\bibliography{sample-base}










\end{document}